\documentclass[onecolumn,letterpaper,11pt,draftclsnofoot]{IEEEtran}

\usepackage{times,amsmath,epsfig}
\usepackage{cite}
\usepackage{graphicx}
\usepackage{psfrag}
\usepackage{subfigure}
\usepackage{url}
\usepackage{stfloats}
\usepackage{amsmath}
\usepackage{amssymb}
\usepackage{array}
\usepackage{array}

\begin{document}

\title{Performance Analysis of Bidirectional Relay Selection with Imperfect Channel State Information}

\author{\IEEEauthorblockN{Hongyu Cui, Rongqing Zhang, Lingyang Song, and Bingli
Jiao}\\
\IEEEauthorblockA{School of Electronics Engineering and Computer
Science\\ Peking University, Beijing, China, $100871$\\
}}

\maketitle
\begin{abstract}
In this paper, we investigate the performance of bidirectional relay
selection using amplify-and-forward protocol with imperfect channel state
information, i.e., delay effect and channel estimation error. The
asymptotic expression of end-to-end SER in high SNR regime is
derived in a closed form, which indicates that the delay effect
causes the loss of both coding gain and diversity order, while the
channel estimation error merely affects the coding gain. Finally,
analytical results are verified by Monte-Carlo simulations.
\end{abstract}

\section{Introduction}

Bidirectional relay communications, in which two sources exchange
information through intermediate relays, have gained a lot of
interest by now, and different transmission schemes have been
proposed \cite{Katti2008}. In \cite{Popovski2007,Louie2010}, an
amplify-and-forward~(AF) based network coding scheme, named as
analog network coding~(ANC), was introduced. With ANC, the data
transmission process can be divided into two phases, and the
spectral efficiency, which is restricted by half-duplex antennas,
can get improved. Recently, relay selection~(RS) for bidirectional
relay networks has been intensively researched to achieve full
spatial diversity and better system performance, which requires
fewer orthogonal resources in comparison of all-participate relay
approaches\cite{Bletsas2006,Ibrahim2008}. Performing RS, the
``best'' relay is firstly selected before data transmission by the
predefined criterion
\cite{Zhang2009,Kyu2009,Song2010,Song2011,Jing2009,Nguyen2010}. In
\cite{Zhang2009,Kyu2009}, the authors proposed the max-min sum rate
selection criterion for AF bidirectional relay. In
\cite{Song2010,Song2011,Jing2009,Nguyen2010}, selection criterions
in minimizing the symbol error rate~(SER) were introduced and
analyzed.

To the authors' best knowledge, most works about RS in bidirectional
relay only consider perfect channel state information~(CSI).
However, imperfect CSI, i.e., delay effect and channel estimation
error~(CEE), has great impact on the performance of bidirectional
relay selection. Specifically, the time delay between relay
selection and data transmission causes that the selected relay may
not be optimal for data
transmission\cite{Torabi2010,Suraweera2010,Michalopoulos2010}. And
similarly, channel estimation errors can not be ignored either
\cite{Seyfi2010,Seung2009,Gedik2009,Ding2011}. In~\cite{Ding2011},
the authors analyzed the performance loss of bidirectional relay
selection using decode-and-forward protocol with CEE, but the impact
of imperfect CSI on a general bidirectional AF relay selection was
not provided.

In light of the aforementioned researches, we analyze the impact of
imperfect CSI, including delay effect and CEE, for bidirectional AF
relay selection in this paper, which has not been studied
previously. The asymptotic expression of end-to-end SER is derived
in a closed form, and verified by computer simulations. Analytical
and simulated results reveal that delay effect reduces both the
diversity order and the coding gain, while channel estimation error
merely causes the coding gain loss. The main contribution of this paper can be summarized as
follows:
\begin{enumerate}
  \item The asymptotic SER expression for bidirectional relay selection is provided in a closed form, which matches the simulated results in high SNR regime;
  \item Imperfect CSI, i.e., delay and channel estimation errors, is taken into account to derive the analytical results, and its therein impact is investigated.
\end{enumerate}

The remainder of this paper is organized as follows: In Section
\uppercase\expandafter{\romannumeral2}, the system model of
bidirectional AF relay selection, and the imperfect CSI model are
described in detail. Section \uppercase\expandafter{\romannumeral3}
provides the analytical expression of bidirectional relay selection
with imperfect CSI. Simulation results and performance analysis are
presented in Section \uppercase\expandafter{\romannumeral4}.
Finally, section \uppercase\expandafter{\romannumeral5} concludes
this paper.

\emph{Notation:}$\left(\cdot\right)^*$ and $\left|\cdot\right|$ represent the conjugate and the absolute value, respectively. $\mathbb{E}$ is used for the expectation and $Pr$ represents the probability. The probability density function and the cumulative probability function of variable $x$ are denoted by $f_x\left(\cdot\right)$ and $F_x\left(\cdot\right)$, respectively.

\section{System Model}

The system investigated in this paper is a general bidirectional AF
relay network with two sources $S_j$, $j=1,2$ exchanging information
through the intermediate $N$ relays $R_i$, $i=1,\ldots,N$. The
direct link between $S_1$ and $S_2$ does not exist, and each node is
equipped with a single half-duplex antenna. The transmit power of
the sources is assumed to be the same, denoted by $p_s$, and all the
relays have the individual power constraint, denoted by $p_r$. The
channel coefficients between sources and relays are reciprocal, and
these coefficients are constant over the duration of one data block.

The whole procedure of bidirectional AF relay selection is divided
into two parts periodically: \emph{relay selection process} and
\emph{data transmission process}, which will be described concretely
in the next section. Let $h_{s,ji}$ and $\hat h_{s,ji}$ represent
the actual and the estimated channel coefficients between $S_j$ and
$R_i$ during the relay selection process, respectively; let
$h_{t,ji}$ and $\hat h_{t,ji}$ represent the actual and the
estimated channel coefficients between $S_j$ and $R_i$ during the
data transmission process, respectively. All the actual channel
coefficients are independent identically distributed~(i.i.d.)
Rayleigh flat-fading with zero mean and unit variance, i.e.,
$\mathbb{E}\left(\left| {h_{t,ji} } \right|^2 \right) =
\mathbb{E}\left(\left| {h_{s,ji} } \right|^2 \right) = 1$, and thus,
$\left| {h_{s,ji} } \right|^2$ and $\left| {h_{t,ji} } \right|^2$
are both exponentially distributed with unit mean. Both the sources
can know the global channel coefficients by estimating the training
symbols, while each relay only has its local channel information.

\subsection{Model of Delay Effect}
Due to the time delay between relay selection process and data
transmission process, $h_{s,ji}$ is not the same as $h_{t,ji}$,
which means the CSI is \emph{outdated}. Their relationship can be
modeled by the first-order autoregressive
model\cite{Michalopoulos2010}:
\begin{equation}\label{Eq:outdate}
h_{t,ji}=\rho_{f_j} h_{s,ji} + \sqrt{1-\rho_{f_j}^2}\varepsilon_j~,
\end{equation}
where $h_{t,ji}$ is a zero mean complex-Gaussian RV with variance of
$\sigma _{h_{t,ji}}^2$; $h_{s,ji}$ and $\varepsilon_j$ are i.i.d.
random variable~(RVs) with zero mean and variance of $\sigma _{
h_{s,ji}}^2$ and $\sigma_{\varepsilon_j}^2$, respectively. In this
paper, we assume $\sigma _{ h_{t,ji}}^2=\sigma _{
h_{s,ji}}^2=\sigma_{\varepsilon_j}^2=1$.

The correlation coefficient $\rho_{f_j}$ ( $0 \le \rho _{f_j } \le
1$, where $\rho_{f_j}=1$ represents no delay effect, in other words,
the CSI is not outdated) between $S_j$ and relays is defined by
Jakes' autocorrelation model \cite{Michalopoulos2010}:
\begin{equation}\label{Eq:rhof}
\rho _{f_j}  = J_0 \left(2\pi f_{d_j} T\right)~,
\end{equation}
where $J_0\left( \cdot \right)$ stands for the zeroth order Bessel
function\cite{Abramowitz}, $f_{d_j}$ is the Doppler frequency, and
$T$ is the time delay between the relay selection process and the
data transmission process. In this paper, two variables
$\rho_{f_j}$, $j=1,2$ are used to represent the correlation
coefficients between $S_j$ and the relays, respectively, for
$f_{d_1}$ and $f_{d_2}$ may be different.

\subsection{Model of Channel Estimation Error}
Let $h$ denote the actual channel coefficient and $\hat h$
represent the estimated channel coefficient, and then their relationship
can be modeled as follows\cite{Seyfi2010}:
\begin{equation}
\hat h = h + e~,\label{Eq:he}
\end{equation}
and
\begin{equation}
h = \rho _e \hat h + d~ \label{Eq:hd}~,
\end{equation}
where $h$ and CEE $e$ are independent complex-Gaussian RVs with zero
mean and variances of $\sigma _h^2$, $\sigma_e^2$, respectively.
$\hat h$ and CEE $d$ are also independent complex-Gaussian RVs with
zero mean and variances of $\sigma_{\hat h}^2$, $\sigma_D^2$,
respectively. The correlation coefficient $ \rho _e  = {\sigma _h^2
}/{ \sigma _{\hat h}^2} $ ~($0 \le \rho_e \le 1$, where $\rho_e =1$
means no CEE)~is determined by the concrete channel estimation
method. In addition, $\rho_e$~can be modeled as an increasing
function of the training symbols' power $P$, i.e., $\rho_e \to 1$
when $P$ approaches infinity\cite{Yoo2006,Feifei2009}. In this
paper, we assume $\rho_e={P}/\left({P+N_0}\right)$\cite{Ramya2009}.

According to the above relationship, the variances of CEE are given
by~:
\begin{equation}\label{Eq:vare}
\sigma _e^2 = \sigma _{\hat h}^2-\sigma _h^2 = \left(1-
\rho_e\right)\sigma _{\hat h}^2 =\frac{1-\rho_e}{\rho_e}\sigma
_h^2~,
\end{equation}
and
\begin{equation}\label{Eq:vard}
\sigma _D^2  = \sigma _h^2  - \rho _e^2 \sigma _{\hat h}^2  = \left(
{1 - \rho _e } \right)\sigma _h^2  = \left( {\rho _e  - \rho _e^2 }
\right)\sigma _{\hat h}^2~.
\end{equation}

Assuming $\sigma_{h}^2 = 1$ in this paper, we have $\sigma_{\hat
h}^2 = \rho_e^{-1}$ and $\sigma_D^2 = 1 - \rho_e$ according
to~(\ref{Eq:vare})~and~(\ref{Eq:vard}).

\subsection{Relationship between~$\hat h_{s,ji}$~and~$\hat h_{t,ji}$~}
For the bidirectional relay selection communications, $\hat
h_{s,ji}$ is used for relay selection, and $\hat h_{t,ji}$ is used
for data detection. According to the model of imperfect CSI, we
have~:

\textbf{\emph{Lemma 1: }}$\hat h_{t,ji}$ and $\hat h_{s,ji}$ can be
related as :
\begin{equation}\label{Eq:lemma1}
\hat h_{t,ji}  = \rho_j \hat h_{s,ji}  + \sqrt {1 - \rho_j ^2 }
v_j~,
\end{equation}
where $v_j$ and $\hat h_{s,ji}$ are i.i.d. RVs, and
\begin{equation}\label{Eq:rho}
\rho_j =
\begin{cases}
  1, ~\text{if } \rho_{f_j}=1; \\
  \rho_e \rho_{f_j}, ~\text{if } \rho_{f_j}<1.
\end{cases}
\end{equation}

When the CSI is not outdated, i.e., $\rho_{f_j}=1$, $\hat \gamma
_{t,ji}=\left| {\hat h_{t,ji} } \right|^2$ and $\hat \gamma
_{s,ji}=\left| {\hat h_{s,ji} } \right|^2$  have the same
distribution.

When the CSI is outdated, i.e., $\rho_{f_j}<1$, the probability
density function~(PDF) of $\hat \gamma _{t,ji}$ conditioned by $\hat
\gamma _{s,ji}$ can be expressed as :
\begin{align}\label{Eq:ccdf}
f_{\hat \gamma _{t,ji} |\hat \gamma _{s,ji} } \left( {y|x} \right) =
\frac{1}{{\left( {1 - \rho _j^2 } \right)\sigma _{\hat h}^2 }}\exp
\left( { - \frac{{x + \rho _j^2 y}}{{\left( {1 - \rho _j^2 }
\right)\sigma _{\hat h}^2 }}} \right)I_0 \left( {\frac{{2\sqrt {\rho
_j^2 xy} }}{{\left( {1 - \rho _j^2 } \right)\sigma _{\hat h}^2 }}}
\right)~,
\end{align}
where $I_0\left(\cdot \right)$ stands for the zeroth order modified
Bessel function of the first kind\cite{Abramowitz}, and $\sigma
_{\hat h}^2  = \sigma _{\hat h_{s,ji} }^2  = \sigma _{\hat h_{t,ji}
}^2  = \rho _e^{ - 1} $.

\textbf{\emph{Proof:}} The proof of Lemma 1 can be found in Appendix
A.$\hfill \blacksquare$

\section{Performance Analysis of Bidirectional Relay Selection with Imperfect CSI}

\subsection{Instantaneous Received SNR at the Sources}

As mentioned above, the whole procedure of bidirectional relay
selection is divided into relay selection process and data
transmission process.

In the relay selection process, the central unit~(CU), i.e., $S_1$
or $S_2$, estimates all the channel coefficients $\hat h_{s,ji}$.
Then, based on the predefined selection criterion, CU selects the
``best" relay from all the available relays for the subsequent data
transmission and other relays keep idle until the next relay
selection instant comes. There are several selection criterions for
bidirectional relay
\cite{Zhang2009,Kyu2009,Song2010,Song2011,Jing2009,Nguyen2010}. In
this paper, we adopt the \emph{Best-Worse-Channel} method for relay
selection which has the best performance in minimizing the average
SER and is tractable for analysis\cite{Jing2009,Nguyen2010}.
According to this criterion, the index $k$ of the selected relay
satisfies~:
\begin{equation}
k = \arg \mathop {\max }\limits_i \min \left\{ \left| {\hat h_{s,1i}
} \right|^2 ,\left| {\hat h_{s,2i} } \right|^2 \right\}~,
\end{equation}
and thus,
\begin{equation}\label{Eq:bestchannel}
 \min \left\{ \left| {\hat h_{s,1k} } \right|^2
,\left| {\hat h_{s,2k} } \right|^2 \right\}  = \mathop {\max
}\limits_i \min \left\{ \left| {\hat h_{s,1i} } \right|^2 ,\left|
{\hat h_{s,2i} } \right|^2 \right\}~.
\end{equation}

The subsequent data transmission process can be divided into two
phases. During the first phase, the sources simultaneously send
their respective information to the intermediate relays where only
the selected relay $R_k$ is active. The superimposed signal at $R_k$
is $ y_k=\sqrt{p_{s}}h_{t,1k}s_1+\sqrt{p_{s}}h_{t,2k}s_2+n_k$, where
$s_j$ denotes the modulated symbols transmitted by $S_j$ with the
average power normalized, $j=1,2$, and $n_k$ is additive white
Gaussian noise~(AWGN) at $R_k$, which is a zero mean
complex-Gaussian RV with two-sided power spectral density of $N_0/2$
per dimension. During the second phase, $R_k$ amplifies the received
signal and forwards it back to the sources. Let $x_k$ be the signal
generated by $R_k$, then we have $x_k = \sqrt {p_r } \beta_k y_k$,
where $\beta_k$ is the amplification factor. In this paper, we
analyze the \emph{variable-gain} AF relay\cite{Seung2009}, then
$\beta_k = \left(p_s \left| {\hat h_{t,1k} } \right|^2 + p_s \left|
{\hat h_{t,2k} } \right|^2 + N_0 \right)^{ -1/2} $ is decided by the
estimated instantaneous channel coefficients.

The received signals by $S_1$ and $S_2$ are similar due to the
symmetry of the network topology, and thus, we take $S_1$ as an
example for analysis. The signal $y_{1,r}$ received by $S_1$ can be
written as $y_{1,r} = h_{t,1k} x_k  + n_1$, where $n_1$ is AWGN at
$S_1$; $n_1$ and $n_k$ are i.i.d. RVs. According to (\ref{Eq:hd}),
$h_{t,1k}$ and $h_{t,2k}$ can be rewritten as $ h_{t,1k}  = \rho _e
\hat h_{t,1k} + d_{1k}$ and $ h_{t,2k}  = \rho _e \hat h_{t,2k}  +
d_{2k}$, where $d_{1k}$ and $d_{2k}$ are independent RVs due to the
independence of $ h_{t,1k} $ and $ h_{t,2k} $. Therefore, $y_{1,r}$
can be expanded as :
\begin{align}
 y_{1,r}  &= \sqrt {p_r p_s } \beta_k \rho _e^2 \hat h_{t,1k} \hat h_{t,2k} s_2  \label{Eq:signal}\\
 &+ \sqrt {p_r p_s } \beta_k \left(\rho _e \hat h_{t,1k} d_{2k}  + \rho _e \hat h_{t,2k} d_{1k}  + d_{1k} d_{2k} \right)s_2  \label{Eq:interI}\\
 &+ \sqrt {p_r p_s } \beta_k \rho _e^2 \hat h_{t,1k} \hat h_{t,1k} s_1\label{Eq:selfI1}\\
 &+ \sqrt {p_r p_s } \beta_k \left(2\rho _e \hat h_{t,1k} d_{1k}  + d_{1k}^2 \right)s_1  \label{Eq:selfI2}\\
 &+ \sqrt {p_r } \beta_k \rho _e \hat h_{t,1k} n_k  + \sqrt {p_r } \beta_k d_{1k} n_k  +
 n_1~,
 \label{Eq:noise}~
\end{align}
where (\ref{Eq:signal}) represents the useful information from
$S_2$; (\ref{Eq:interI}) represents the inter-interference from
$S_2$ caused by CEE; (\ref{Eq:selfI1}) and (\ref{Eq:selfI2})
represent the self-interference from $S_1$ itself which can be
subtracted totally by self-canceling if CEE does not
exist\cite{Song2011}. However, with CEE, $S_1$ can only reconstruct
$\sqrt {p_r p_s } \beta_k \rho _e^2 \hat h_{t,1k} \hat h_{t,1k} s_1$
at the receiver. Thus, only (\ref{Eq:selfI1}) can be subtracted
totally, whereas the self-interference of (\ref{Eq:selfI2}) is
residual; (\ref{Eq:noise}) includes the amplified noise from $R_k$
and the noise at $S_1$.

After self-canceling $\sqrt {p_r p_s } \beta_k \rho _e^2 \hat
h_{t,1k} \hat h_{t,1k} s_1$ from $y_{1,r}$, and then multiplied by
$\hat h_{t,1k}^* \hat h_{t,2k}^* $ to compensate the phase rotation,
the processed signal $y_1$ at $S_1$ is :
\begin{align}\label{Eq:y1}
y_1&=\hat h_{t,1k}^* \hat h_{t,2k}^* \left( y_{1,r} - \sqrt {p_r p_s
} \beta_k \rho _e^2 \hat h_{t,1k} \hat h_{t,1k} s_1\right)~.
\end{align}

The transmitted information $s_2$ can be recovered by maximum
likelihood detection:
\begin{equation}
\tilde s_2  = \arg \mathop {\min }\limits_{ {s'_2  \in \mathcal{A}}
} \left\| {y_1  - \sqrt {p_r p_s } \beta \rho _e^2 \left| {\hat
h_{t,1k} } \right|^2 \left| {\hat h_{t,2k} } \right|^2 s'_2 }
\right\|^2 ~,
\end{equation}
where $ \left\| {\cdot} \right\|^2 $ represents the Euclid-distance,
$ {\mathcal{A}} $ is the alphabet of modulation symbols, and $\tilde
s_2$ is the recovered signal.

According to (\ref{Eq:y1}), the instantaneous received SNR
$\gamma_1$ at $S_1$ can be written as :
\begin{equation}\label{Eq:gamma1}
\gamma _1  = \frac{{\psi _r \psi _s \rho _e^4 \left| {\hat h_{t,1k}
} \right|^2 \left| {\hat h_{t,2k} } \right|^2 }}{{\left(5\psi _r
\psi _s \rho _e^2 \sigma _D^2  + \psi _r \rho _e^2  + \psi _s
\right)\left| {\hat h_{t,1k} } \right|^2  + \left(\psi _r \psi _s
\rho _e^2 \sigma _D^2 + \psi _s \right)\left| {\hat h_{t,2k} }
\right|^2 + 3\psi _r \psi _s \sigma _D^4  + \psi _r \sigma _D^2  +
1}}~,
\end{equation}
where $ \psi _s = p_s / N_0 $, $ \psi _r = p_r / N_0 $, $\rho_e$ is
the CEE coefficient, and the CEE variance $\sigma_D^2=1-\rho_e$.

In high SNR regime, $\rho_e \to 1$ and
$\sigma_D^2=\left(1-\rho_e\right) \to 0$, then the item $3\psi _r
\psi _s \sigma _D^4 + \psi _r \sigma _D^2 + 1$ in the denominator of
(\ref{Eq:gamma1}) approaches 1, which can also be ignored when SNR
approaches infinity\cite{Song2011}.

Therefore, $\gamma_1$ in high SNR regime can be simplified into~:
\begin{equation}\label{Eq:gamma1high}
\gamma _1  = \frac{{\tilde a \left| {\hat h_{t,1k} } \right|^2
\tilde b \left| {\hat h_{t,2k} } \right|^2 }}{{\tilde a \left| {\hat
h_{t,1k} } \right|^2 + \tilde b \left| {\hat h_{t,2k} } \right|^2
}}~,
\end{equation}
 where
\begin{equation}
\tilde a  = \frac{{\psi_r  \rho _e^4 }}{{1 + \psi_r
 \rho _e^2 \sigma _D^2 }}~,~
\tilde b  = \frac{{\psi_r \psi_s \rho _e^4  }}{{5\psi_r \psi_s \rho
_e^2 \sigma _D^2  + \psi_r \rho _e^2   + \psi_s  }}~.
\end{equation}

$\gamma_1$ in (\ref{Eq:gamma1high}) is greater than that in
(\ref{Eq:gamma1}), whereas they match tightly in high SNR regime.
Therefore, we use $\gamma_1$ in (\ref{Eq:gamma1high}) for asymptotic
analysis in the followings.

\subsection{Distribution Function of the Received SNR}

The distribution of $\gamma_1$ in (\ref{Eq:gamma1high}) is decided
by $\hat \gamma_{t,1k}=\left| {\hat h_{t,1k} } \right|^2$ and $\hat
\gamma_{t,2k}=\left| {\hat h_{t,2k} } \right|^2$, which are
determined by $\hat \gamma_{s,1k}=\left| {\hat h_{s,1k} } \right|^2$
and $\hat \gamma_{s,2k}=\left| {\hat h_{s,2k} } \right|^2$ according
to Lemma 1. Furthermore, the distribution of $\hat \gamma_{s,1k}$
and $\hat \gamma_{s,2k}$ can be obtained by the above selection
criterion. After some manipulations, we have

\textbf{\emph{Theorem 1:}} With the definition that~:
\begin{align}\label{Eq:a}
a \buildrel \Delta \over = \frac{{\rho _e }}{{\tilde a}} = \frac{{1
+ \psi _r \rho _e^2 \sigma _D^2 }}{{\psi _r \rho _e^3 }}~,~ b
\buildrel \Delta \over = \frac{{\rho _e }}{{\tilde b}} =
\frac{{5\psi _r \psi _s \rho _e^2 \sigma _D^2  + \psi _r \rho _e^2 +
\psi _s }}{{\psi _r \psi _s \rho _e^3 }}~,
\end {align}
the cumulative distribution function~(CDF) of  $\gamma_1$ is~:
\begin{align}\label{Eq:cdf}
F_{\gamma _1 } \left(z\right) = 1 - N^2\sum\limits_{m = 0}^{N - 1}
{\sum\limits_{n = 0}^{N - 1} \binom{N-1}{m}} \binom{N-1}{n}
\frac{\left(-1\right)^m}{{2m + 1}}\frac{\left(-1\right)^n}{{2n + 1}}
\left(f_1 + f_2 +f_3 + f_4 \right) ~
\end{align}
where
\begin{equation}
f_1  = 2\sqrt {ab} z\exp \left( - \left(a + b\right)z\right)K_1
\left(2z\sqrt {ab} \right)~,
\end{equation}
\begin{align}
f_2 &= \sqrt {\frac{{16n^2 ab}}{{2\left(n + 1\right)\left[\left(2n +
1\right)\left(1 - \rho _1 \right) + 1\right]}}} z\exp \left( -
\left(\frac{{2\left(n + 1\right)a}}{{\left(2n + 1\right)\left(1 -
\rho _1 \right) + 1}}+b\right)z\right)\\\notag &\times K_1
\left(2z\sqrt {\frac{{2\left(n + 1\right)ab}}{{\left(2n +
1\right)\left(1 - \rho _1 \right) + 1}}}\right)~,
\end{align}
\begin{align}
f_3  &= \sqrt {\frac{{8m^2 ab}}{{\left(m + 1\right)\left[\left(2m +
1\right)\left(1 - \rho _2 \right) + 1\right]}}} z\exp \left( -
\left(a + \frac{{2\left(m + 1\right)b}}{{\left(2m + 1\right)\left(1
- \rho _2 \right) + 1}}\right)z\right)\\\notag &\times K_1
\left(2z\sqrt {\frac{{2ab\left(m + 1\right)}}{{\left(2m +
1\right)\left(1 - \rho _2 \right) + 1}}}\right)~,
\end{align}
\begin{align}
f_4  &= \sqrt {\frac{{16m^2 n^2 ab}}{{\left(m + 1\right)\left(n
+1\right)\left[\left(2n + 1\right)\left(1 -\rho _1 \right) +
1\right]\left[\left(2m + 1\right)\left(1 - \rho _2 \right)+
1\right]}}} z\\\notag &\times \exp \left( - \left(\frac{{2\left(n +
1\right)a}}{{\left(2n + 1\right)\left(1 -\rho _1 \right) + 1}} +
\frac{{2\left(m + 1\right)b}}{{\left(2m + 1\right)\left(1 - \rho _2 \right) + 1}}\right)z\right)\notag\\
&\times K_1 \left( 4z\sqrt {\frac{{ab\left(m + 1\right)\left(n +
1\right)}}{{\left[\left(2m +1\right)\left(1 - \rho _2 \right) +
1\right]\left[\left(2n+ 1\right)\left(1 - \rho _1 \right) +
1\right]}}} \right)~.
\end{align}
And $K_1\left(\cdot\right)$ is the first order modified Bessel of
the second kind\cite{Abramowitz}, $\binom{N}{k}$ is the binomial
coefficient, and $\rho_j$, $j=1,2$ satifies (\ref{Eq:rho}) in Lemma
1: $\rho_j=1$ if $\rho_{f_j}=1$, and $\rho_j=\rho_e\rho_{f_j}$ if
$\rho_{f_j}<1$.

\emph{\textbf{Proof: }}The proof of Theorem 1 can be found in
Appendix B. $\hfill \blacksquare$

Due to the symmetry, it can be proved similarly that the CDF of the
received SNR $\gamma_2$ at $S_2$ have the same form as $\gamma_1$,
and their PDFs can be obtained by differentiating the CDFs.

\subsection{Asymptotic Performance of Average Symbol Error Rate}

For many common modulation formats, the average SER can be obtained
by\cite{Suraweera2010}:
\begin{align}
\overline {SER}  = \alpha \mathbb{E}  \left[Q\left(\sqrt {\beta
\gamma } \right)\right]= \frac{\alpha }{{\sqrt {2\pi }
}}\int\limits_0^\infty {F_\gamma \left(\frac{{t^2 }}{\beta
}\right)e^{ - \frac{{t^2 }}{2}} } dt~,\label{Eq:ps2}
\end{align}
where $\gamma$ is the instantaneous received SNR,
$Q\left(\cdot\right)$ is Gaussian Q-Function\cite{Abramowitz}, and
$\alpha=1$,~$\beta=2$ for BPSK, $\alpha=1$,~$\beta=1$ for QPSK,
$\alpha=1/\log _2 M$,~$\beta=\log _2 M\sin ^2 (\pi /M)$ for
MPSK~($M>4$).

Applying Theorem 1 and (\ref{Eq:ps2}), the exact average SER of
$S_1$ can be obtained by \cite[(6.621.3)]{Gradshteyn94}:
\begin{align}
\int_0^\infty {x^{\mu  - 1} } e^{ - \alpha x} K_\nu  \left(\beta
x\right)dx = \frac{{\sqrt \pi \left(2\beta \right)^\nu
}}{{\left(\alpha  + \beta \right)^{\mu + \nu } }}\frac{{\Gamma
\left(\mu  + \nu \right)\Gamma \left(\mu  + \nu \right)}}{{\Gamma
\left(\mu + 1/2\right)}}F\left(\mu  + \nu ,\nu + \frac{1}{2};\mu +
\frac{1}{2};\frac{{\alpha  - \beta }}{{\alpha + \beta }}\right)~,
\end{align}
where $\Gamma\left(\cdot\right)$ is Gamma function, and
$F\left(\cdot\right)$ is Confluent Hypergeometric
function\cite{Abramowitz}. However, the exact form is too
complicated to analyze the performance, thus we resort to the high
SNR analysis\cite{Zheng2003}.


\emph{\textbf{Theorem 2:}} The asymptotic performance of SER in high
SNR regime can be obtained in two different cases according to
whether the CSI is outdated or not.

\begin{itemize}
  \item When the CSI is not outdated, i.e., the delay coefficients satisfy
$\rho_{f_1}=\rho_{f_2} =1$~, and the CEE coefficient $\rho_e$ is
arbitrary, the average SER of $S_1$ in high SNR regime is:
\begin{align}\label{Eq:SER11}
\overline {SER}_1^\infty   = \frac{\alpha }{{4\beta ^N
}}\frac{{\left(2N\right)!}}{{N!}}\left(\left(\frac{{1 + \psi _r \rho
_e^2 \sigma _D^2 }}{{\psi _r \rho _e^3 }}\right)^N  +
\left(\frac{{5\psi _r \psi _s \rho _e^2 \sigma _D^2  + \psi _r \rho
_e^2 + \psi _s }}{{\psi _r \psi _s \rho _e^3 }}\right)^N \right)~,
\end{align}
where $\alpha$ and $\beta$ are decided by the modulation format in
(\ref{Eq:ps2}); $\psi_s=p_s/N_0$ and $\psi_r=p_r/N_0$; $N!$ is the
factorial of $N$; $\sigma_D^2=1-\rho_e$.

  \item When the CSI is outdated, i.e., $ \rho _{f_1} < 1$~or~$\rho_{f_2}
<1$, and $\rho_e$ is arbitrary, the average SER of $S_1$ in high SNR
regime is:
\begin{align}\label{Eq:SER12}
\overline {SER}_1^\infty = &\frac{\alpha }{{2\beta }}\left(\frac{{1
+ \psi _r \rho _e^2 \sigma _D^2 }}{{\psi _r \rho _e^3
}}\right)N\sum\limits_{n = 0}^{N - 1} {\left( - 1\right)^n }
\binom{N-1}{n}\frac{{2 - \rho _1 }}{{\left(2n + 1\right)\left(1 -
\rho _1 \right) + 1}} \\\notag+ &\frac{\alpha }{{2\beta
}}\left(\frac{{5\psi _r \psi _s \rho _e^2 \sigma _D^2  + \psi _r
\rho _e^2 + \psi _s }}{{\psi _r \psi _s \rho _e^3
}}\right)N\sum\limits_{m = 0}^{N - 1} {\left( - 1\right)^m }
\binom{N-1}{m}\frac{{2 -\rho _2 }}{{\left(2m+ 1\right)\left(1 - \rho
_2 \right) +1}}~,
\end{align}
where $\rho_j$, $j=1,2$ satifies (\ref{Eq:rho}) in Lemma 1:
$\rho_j=1$ if $\rho_{f_j}=1$, and $\rho_j=\rho_e\rho_{f_j}$ if
$\rho_{f_j}<1$.
\end{itemize}

 \emph{\textbf{Proof: }}The proof of Theorem 2 can be found in
Appendix C.$\hfill \blacksquare$

 Similarly, the average SER in high SNR regime of $S_2$ can be obtained from
(\ref{Eq:SER11}) and (\ref{Eq:SER12}) by permuting $a$ with $b$.

\subsection{Performance Analysis of Diversity Order and Coding Gain}

Diversity order $d =  - \mathop {\lim }\limits_{\psi _t \to \infty }
\big( {\log \overline {SER} _1^\infty  /\log \psi _t } \big) $
\cite{Zheng2003}, where
$\psi_t=\left(2p_s+p_r\right)/N_0=2\psi_s+\psi_r$, is an useful
metric to describe the asymptotic performance of SER, i.e., greater
diversity order means the curve of SER attenuates more quickly.

\emph{\textbf{Theorem 3:}} According to the definition of diversity
order, the diversity order is :
\begin{equation}
d =
\begin{cases}
  N, ~\text{if the CSI is not outdated, i.e.,~} \rho_{f_1}=\rho_{f_2}=1; \\
  1, ~\text{if the CSI is outdated, i.e.,~} \rho_{f_1}<1 \text{~or~}\rho_{f_2}<1.
\end{cases}
\end{equation}

 \emph{\textbf{Proof: }} Assuming $\psi_t=\left(2p_s+p_r\right)/N_0=2\psi_s+\psi_r$,
$\psi_s=p_s/N_0=\lambda \psi_t$, and
$\psi_r=p_r/N_0=\left(1-2\lambda\right) \psi_t$, the diversity order
can be obtained by Theorem 2, and the fact that $\rho_e \to 1$ and
$\sigma_D^2=1-\rho_e \to 0$ when SNR approaches infinity.$\hfill
\blacksquare$

Theorem 3 reveals that the diversity order is $N$ if and only if the
CSI is not outdated. Once the CSI is outdated, i.e., the delay
exists, the diversity order reduces to $1$, whereas CEE has no
impact on the performance loss of diversity order.

However, both delay effect and CEE can reduce the coding gain, which
is the shift of SER curve, e.g., different delay
coefficients $\rho_{f_j}$ and CEE coefficients $\rho_e$ will result
in different $\rho_j$ is Theorem 2, and thus the coding gain is
different.

\section{Simulation Results and Discussion}

In this section, the average SER of bidirectional relay selection
with imperfect CSI is studied by Monte-Carlo simulations, and the
analytical performance provided by Theorem 2 is verified by these
simulation results. Due to the symmetry of the network, the
following results only concern about the average SER of $S_1$. All
the simulations are performed with BPSK modulation over the
normalized Rayleigh fading channels. For simplicity, we assume that
sources and relays have the same power, i.e., $p_s=p_r=P_0$, and the
x-axis of the following figures is $\mbox{SNR}=P_0/N_0$ in decibel.
To better understand the impact of imperfect CSI, we discuss four
different situation, i.e., perfect CSI, only delay effect, only CEE,
and both delay effect and CEE.

In Fig. 1, we compare the simulated and the analytical SER of
bidirectional relay selection with perfect CSI for $N$ relays, i.e.,
$\rho_{f_1}=\rho_{f_2}=1$ and $\rho_e=1$. This figure shows that
increasing the number of available relays can reduce the average
SER, because the diversity order is $N$ when the CSI is perfect.
This figure also shows that the asymptotic analytical SER given by
Theorem 2 is the lower bound of the simulated results due to the
fact that $\gamma_1$ in (\ref{Eq:gamma1high}) is greater than that
in (\ref{Eq:gamma1}), whereas both the analytical and the simulated
results match tightly in high SNR regime.

In Fig. 2, we analyze the impact of delay on the SER performance
without CEE~, i.e., $\rho_e=1$. For simplicity, we assume
$\rho_{f_1}=\rho_{f_2}=\rho_f$ and $N=4$. The figure reveals that
the diversity order degrades to 1 once $\rho_f<1$ regardless of $N$.
Although the diversity order is 1 once $\rho_f<1$, yet the coding
gain is different for different $\rho_f$. Comparing the curves of
$\rho_f=0.9$ and $\rho_f=0$, the coding gain gap between them is
approximately 6dB in high SNR regime. Besides, the performance at
moderate SNR is different for different $\rho_f$, i.e., greater
$\rho_f$ has better performance at moderate SNR. For example, at
moderate SNR, i.e., range from 8dB to 16dB, the slope of the SER
curve of $\rho_f=0.9$ is greater than $1$, while the slope of
$\rho_f=0$ at the same range is $1$. The performance at moderate SNR
can be analyzed by the exact expression of SER and Maclaurian
Series\cite{Zheng2003}.

In Fig. 3, we study the impact of CEE on the SER performance without
delay, i.e., $\rho_f=1$ and $\rho_e={P}/\left({P+N_0}\right)$, where
$P$ is the power of the training symbols\cite{Ramya2009}. $P$ can be
greater than the power of the data symbols $P_0$ to obtain better
performance of channel estimation, thus we simulate the situation of
$P=P_0, 2P_0, 4P_0$ and $\infty$ ($P=\infty$ means no CEE),
respectively. With CEE, the diversity order is invariant, which is
the same as the number of relays. However, compared with the curve
of $P=\infty$, there exists coding gain loss caused by CEE, and the
loss could be reduced by increasing the power of training symbols
$P_0$. As Fig. 3 illustrated, the coding gain loss in high SNR
regime is about 5dB when $P=P_0$, but it reduces to 2dB when
$P=4P_0$.

In Fig. 4, the joint effect of delay and channel estimation error is
considered and compared with the cases of only delay effect, only
CEE, and perfect CSI. The results also indicate that delay will result
in the diversity order loss and the coding gain loss, and CEE will
merely result in the coding gain loss. With both delay and CEE
existing, the SER performance is the worst, which matches tightly
with the analytical result in high SNR regime.

\section{Conclusions}

In this paper, we analyzed the performance of bidirectional AF relay
selection with imperfect CSI, i.e., delay effect and channel
estimation error, and the asymptotic analytical expression of
end-to-end SER was derived and verified by the computer simulation.
Both analytical and simulated results indicate that delay effect results in
the coding gain loss and the diversity order loss, and channel
estimation error will merely cause the coding gain loss.

\section*{Appendix A\\Proof of Lemma 1}
At the case of $\rho_{f_j}=1$, we have $h_{t,ji} = h_{s,ji}$ by
(\ref{Eq:outdate}), and thus $\hat h_{t,ji}=\hat h_{s,ji}$, which is a
special case of (\ref{Eq:lemma1}) when $\rho_j=1$ in Lemma 1.

At the case of $\rho_{f_j}<1$, by (\ref{Eq:outdate}), (\ref{Eq:he})
and (\ref{Eq:hd}), we have~:
\begin{align}
\hat h_{t,ji}= h_{t,ji}+e=\rho_{f_j}
h_{s,ji}+\sqrt{1-\rho_{f_j}^2}\varepsilon_j+e=\rho_{f_j}\rho_e \hat
h_{s,ji}+\rho_{f_j}d+\sqrt{1-\rho_{f_j}^2}\varepsilon_j+e~,
\end{align}
where $d$, $\varepsilon_j$, and $e$ are independent zero mean
complex-Gaussian RVs with variance of $\sigma_D^2$,
$\sigma_{\varepsilon_j}^2$, and $\sigma_e^2$, respectively. Thus,
$\rho_{f_j}d+\sqrt{1-\rho_{f_j}^2}\varepsilon_j+e$ is a zero mean
complex-Gaussian RV with variance of
$\rho_{f_j}^2\sigma_D^2+\left(1-\rho_{f_j}^2\right)\sigma_{\varepsilon_j}^2+\sigma_e^2$~,
which can be simplified into
$\left(1-\rho_{f_j}^2\rho_e^2\right)\sigma_{\hat h_{t,ji}}^2$ by the
relationship of variances (\ref{Eq:vare}),(\ref{Eq:vard}). Then,
$\rho_{f_j}d+\sqrt{1-\rho_{f_j}^2}\varepsilon_j+e$ can be written as
$\sqrt{1-\rho_{f_j}^2\rho_e^2}v_j$~, where $v_j$ is an independent
RV with zero mean and variance of $\sigma_{\hat h_{t,ji}}^2$.
Defining $\rho_j=\rho_e\rho_{f_j}$ , formula (\ref{Eq:lemma1}) in
Lemma 1 is proved. Thus, $\hat h_{t,ji}$ and $\hat h_{s,ji}$ are
jointly complex-Gaussian, and $\hat \gamma_{t,ji}=\left|\hat
h_{t,ji}\right|^2$ and $\hat \gamma_{s,ji}=\left|\hat
h_{s,ji}\right|^2 $ are correlated exponential distributions, then
the joint PDF $f_{\hat \gamma_{t,ji} ,\hat \gamma_{s,ji} } \left(
{y,x} \right)$ is given by\cite{Simon}:
\begin{align}
f_{\hat \gamma_{t,ji} ,\hat \gamma_{s,ji} } \left( {y,x} \right) =
\frac{1}{{\left( {1 - \rho _j^2 } \right)\sigma _{\hat h}^4 }}\exp
\left( { - \frac{{x + y}}{{\left( {1 - \rho _j^2 } \right)\sigma
_{\hat h}^2 }}} \right)I_0 \left( {\frac{{2\sqrt {\rho _j^2 xy}
}}{{\left( {1 - \rho _j^2 } \right)\sigma _{\hat h}^2 }}} \right)~.
\end{align}
And now, the conditional probability of (\ref{Eq:ccdf}) in Lemma 1
can be proved by\cite{Paoulis}:
\begin{align}
 f_{\hat \gamma_{t,ji}|\hat
\gamma_{s,ji}} \left( {y|x} \right)=\frac{f_{\hat \gamma_{t,ji}
,\hat \gamma_{s,ji}} \left( {y,x} \right)}{f_{\hat \gamma_{s,ji} }
\left( x \right)}~,
\end{align}
where $f_{\hat \gamma _{s,ji} } \left( x \right) = \exp \left( { -
x/\sigma _{\hat h}^2 } \right)/\sigma _{\hat h}^2 $~.

\section*{Appendix B\\ Proof of Theorem 1}

\subsection{distribution of $\hat \gamma_{s,1k} $ and $\hat \gamma_{s,2k} $}
Following the similar steps of \cite{Michalopoulos2010} , the CDF of
$\hat \gamma _{s,1k}$ can be expressed as :
\begin{align}\label{Eq:T10}
F_{\hat \gamma _{s,1k} } \left( x \right) & \buildrel \left(a\right)
\over= N\Pr \left\{ {\hat \gamma _{s,1i}  < x,k = i} \right\}
\\\notag
& \buildrel \left(b\right) \over = N\int\limits_0^x {f_{\hat \gamma
_{s,1i} } \left(y\right)} \Pr \left\{ {\hat \gamma _{s,1i}  \le \hat
\gamma _{s,2i} |\hat \gamma _{s,1i}  = y} \right\}\Pr \left\{ {k =
i|\hat \gamma _{s,1i}  \le \hat \gamma _{s,2i} ,\hat \gamma _{s,1i}
= y} \right\}dy
\\\notag
&+ N\int\limits_0^x {f_{\hat \gamma _{s,1i} } (y)} \Pr \left\{ {\hat
\gamma _{s,1i}  > \hat \gamma _{s,2i} |\hat \gamma _{s,1i}  = y}
\right\}\Pr \left\{ {k = i|\hat \gamma _{s,1i}  > \hat \gamma
_{s,2i} ,\hat \gamma _{s,1i}  = y} \right\}dy~
\end{align}
where (a) in (\ref{Eq:T10}) is satisfied due to the symmetry among the $N$ end-to-end
paths, and (b) is satisfied by dividing the union event into two
disjoint events, i.e., $\hat \gamma _{s,1i}  > \hat \gamma _{s,2i}$
and $\hat \gamma _{s,1i}  \le \hat \gamma _{s,2i}$. According to the
selection criterion (\ref{Eq:bestchannel}) and order statistics of
independent RVs\cite{David1970}: $ \Pr \big\{ {\min \big( {x1,x2}
\big) \le z} \big\} = 1 -
\left(1-F_{x_1}\left(z\right)\right)\left(1-F_{x_2}\left(z\right)\right)
$, and the fact that $ F_{\hat \gamma _{s,1i} } \left( z \right) =
F_{\hat \gamma _{s,2i} } \left( z \right) = 1 - \exp \left( { -
z/\sigma _{\hat h}^2 } \right) $, we have
\begin{align}\label{Eq:T11}
\Pr \left\{ {k = i|\hat \gamma _{s,1i}  \le \hat \gamma _{s,2i}
,\hat \gamma _{s,1i}  = y} \right\} = \prod\limits_{p \ne i} {\Pr }
\left\{ {\min \left( {\hat \gamma _{s,1i} ,\hat \gamma _{s,2i} }
\right) \le y} \right\} = \left( {1 - \exp \left( { -
\frac{2y}{\sigma _{\hat h}^2 } }\right)} \right)^{N - 1}~.
\end{align}
Similarly, the conditional probability $ \Pr \left\{ {k = i|\hat
\gamma _{s,1i}  > \hat \gamma _{s,2i} ,\hat \gamma _{s,1i}  = y}
\right\} $ can be achieved. Therefore, substituting (\ref{Eq:T11})
into (\ref{Eq:T10}), $F_{\hat \gamma _{s,1k} } \left( x \right)$ can
be written as :
\begin{align}
F_{\hat \gamma _{s,1k} } \left( x \right) &=~N\int_0^x
{\frac{1}{{\sigma _{\hat h}^2 }}\exp \left( { - \frac{y}{{\sigma
_{\hat h}^2 }}} \right)} \left( {\int_0^y {\frac{1}{{\sigma _{\hat
h}^2 }}\exp \left( { - \frac{z}{{\sigma _{\hat h}^2 }}} \right)}
\left( {1 - \exp \left( { - \frac{{2z}}{{\sigma _{\hat h}^2 }}}
\right)} \right)^{N - 1} dz} \right)dy
\\\notag
&+~N\int_0^x {\frac{1}{{\sigma _{\hat h}^2 }}\exp \left( { -
\frac{y}{{\sigma _{\hat h}^2 }}} \right)} \left( {\int_y^\infty
{\frac{1}{{\sigma _{\hat h}^2 }}\exp \left( { - \frac{z}{{\sigma
_{\hat h}^2 }}} \right)} dz} \right)\left( {1 - \exp \left( { -
\frac{{2y}}{{\sigma _{\hat h}^2 }}} \right)} \right)^{N - 1} dy~.
\end{align}
Applying binomial expansion $\left( {1 - x} \right)^N  =
\sum\nolimits_{k = 0}^N {\binom{N}{k}} \left( { - 1} \right)^k x^k$
 and $
N\sum\nolimits_{n = 0}^{N - 1} \binom{N-1}{n} {{\left( { - 1}
\right)^n }}/\left({{n + 1}}\right) = 1 $ in
\cite[(0.155.1)]{Gradshteyn94}, $F_{\hat \gamma _{s,1k} } \left( x
\right)$~can be rewritten as :
\begin{align}\label{Eq:Fs1k}
F_{\hat \gamma _{s,1k} } \left( x \right) = 1 - &N\sum\limits_{n =
0}^{N - 1} \binom{N-1}{n} \frac{{\left( { - 1} \right)^n }}{{2n +
1}}\left[\exp \left( { - \frac{x}{{\sigma _{\hat h}^2 }}} \right) +
\frac{n}{{n + 1}}\exp \left( { - \frac{{2\left( {n + 1}
\right)x}}{{\sigma _{\hat h}^2 }}} \right)\right]~,
\end{align}
where $\sigma_{\hat h}^2=\rho_e^{-1}$, and it can be proved
similarly that the CDF of~$\hat \gamma_{s,1k}$~have the same form,
and their PDFs can be obtained by differentiating the CDFs.

\subsection{distribution of $\gamma_1$}

At the case of $\rho_{f_1} < 1$ and $\rho_{f_2}<1$, and by Lemma 1
and $ \int\limits_0^\infty {\exp \left( - \alpha x\right)I_0
\left(\beta \sqrt x \right)} dx = \left({1}/{\alpha }\right)\exp
\left({{\beta ^2 }}/\left({{4\alpha }}\right)\right)$ in
\cite[(6.614.3)]{Gradshteyn94}, we have~:
\begin{align}
f_{\hat \gamma _{t,1k} } \left( x \right) &= N\sum\limits_{n = 0}^{N
- 1} {\left( { - 1} \right)^n \binom{N-1}{n}} \frac{1}{{2n +
1}}\left[ \frac{{\exp \left( { - x/\sigma _{\hat h}^2 }
\right)}}{{\sigma _{\hat h}^2 }}
\right.\\\notag&\left.+\frac{{2n/\sigma _{\hat h}^2 }}{{\left( {2n +
1} \right)\left( {1 - \rho_1^2 } \right) + 1}}\exp \left( { -
\frac{{2 \left( {n + 1} \right)x/\sigma _{\hat h}^2 }}{{\left( {2n +
1} \right)\left( {1 - \rho_1^2 } \right) + 1}}} \right) \right]~.
\end{align}
The CDF of $\hat \gamma_{t,1k}$ can be obtained by integrating the
PDF, and the distribution of $\hat \gamma _{t,2k}$ can be obtained
by substituting $\rho_1$ with $\rho_2$. Let $\Omega_1$ and
$\Omega_2$ represent $\tilde a\hat \gamma _{t,1k}$, and $\tilde b
\hat \gamma _{t,2k} $ respectively, and the distribution of
$\Omega_1$ and $\Omega_2$ can be obtained by $f_Y \left(y\right) =
\left({1}/{m}\right)f_X \left({x}/{m}\right)$, and $F_Y \left( y
\right) = F_X \left( {x}/{m} \right) $  when $ Y = mX \left(m
> 0\right) $\cite{Paoulis}. Thus, the CDF of $\gamma_1$ can be written as
:
\begin{align}\label{Eq:FF}
F_{\gamma _1 } \left( z \right) &= \Pr \left\{ {\frac{{\Omega _1
\Omega _2 }}{{\Omega _1  + \Omega _2 }} < z} \right\}
\\\notag
& = \Pr \left\{ {\left( {\Omega _2  - z} \right)\Omega _1  < z\Omega
_2 ,\Omega _2  > z} \right\} + \Pr \left\{ {\left( {\Omega _2  - z}
\right)\Omega _1  < z\Omega _2 ,\Omega _2  \le z} \right\}
\\\notag
&= \int_z^\infty  {F_{\Omega _1 } \left( {\frac{{zx}}{{x - z}}}
\right)} f_{\Omega _2 } \left(x\right)dx + \int_0^z {\left[ {1 -
F_{\Omega _1 } \left( {\frac{{zx}}{{x - z}}} \right)} \right]}
f_{\Omega _2 } \left( x \right)dx
\\\notag
&= 1 - \int\limits_0^\infty  {f_{\Omega _2 } \left( {x + z}
\right)\left[ {1 - F_{\Omega _1 } \left( {z + \frac{{z^2 }}{x}}
\right)} \right]} dx
\end{align}
Substituting $ \int\limits_0^\infty  {\exp \left( - mx - nx^{ - 1}
\right)} dx = 2\sqrt {{n}/{m}} K_1 \left(2\sqrt {mn} \right)$ in
\cite[(3.324)]{Gradshteyn94} into (\ref{Eq:FF}), Theorem 2 can be
proved when using $N\sum\nolimits_{n = 0}^{N - 1}
{\binom{N-1}{n}{{\left( { - 1} \right)^n }}/\left({{n + 1}}\right)}
= 1 $ in \cite[(0.155.1)]{Gradshteyn94}.

At the case of $\rho_{f_1}=1$ or $\rho_{f_2}=1$, it can be proved in
a similar way that the CDF of $\gamma_1$ can also be expressed as
the formula (\ref{Eq:cdf})~in Theorem 1.

\section*{Appendix C\\ Proof of Theorem 2 }

In high SNR regime~\big($\psi_r$,$\psi_s$ $\to$ $\infty$\big),
$\rho_e \to 1$ and $a,b \to 0$. By applying the Bessel function
approximation for small $x \to 0$, $K_1 \left( x \right) \approx
{1}/{x} $\cite{Abramowitz} in Theorem 1, we have~:
\begin{align}\label{Eq:hignFgamma}
F_{\gamma _1 }  \left( z \right) &\approx 1 - N^2 \sum\limits_{m =
0}^{N - 1} {\sum\limits_{n = 0}^{N - 1} {\left( { - 1} \right)^{m +
n} \binom{N-1}{m}} } \binom{N-1}{n}\frac{1}{{2m + 1}}\frac{1}{{2n +
1}}
\\\notag
&\times \bigg[\exp \left( { - \left( {a + b} \right)z} \right) +
\frac{n}{{n + 1}}\exp \left( { - \left( {\frac{{2\left( {n + 1}
\right)a}}{{\left( {2n + 1} \right)\left( {1 - \rho _1 } \right) +
1}} + b} \right)z} \right) \\\notag
 &+ \frac{m}{{m + 1}}\exp
\left( { - \left( {a + \frac{{2\left( {m + 1} \right)b}}{{\left( {2m
+ 1} \right)\left( {1 - \rho _2 } \right) + 1}}} \right)z}
\right)\\\notag
 &+ \frac{m}{{m + 1}}\frac{n}{{n + 1}}\exp
\left( { - \left( {\frac{{2\left( {n + 1} \right)a}}{{\left( {2n +
1} \right)\left( {1 - \rho _1 } \right) + 1}} + \frac{{2\left( {m +
1} \right)b}}{{\left( {2m + 1} \right)\left( {1 - \rho _2 } \right)
+ 1}}} \right)z} \right)\bigg]~.
\end{align}

At the case of $\rho_{f_1}=\rho_{f_2}=1$, $F_{\gamma _1 } \left( z
\right)$ can be rewritten as :
\begin{align}
F_{\gamma _1 } \left( z \right) &= 1 - N\sum\limits_{n = 0}^{N - 1}
{\left( { - 1} \right)^n } \binom{N-1}{n}\frac{1}{{2n + 1}}\left[
{\exp \left( { - az} \right) + \frac{n}{{n + 1}}\exp \left( { -
2\left( {n + 1} \right)az} \right)} \right]\\\notag &\times
N\sum\limits_{m = 0}^{N - 1} {\left( { - 1} \right)^m }
\binom{N-1}{m}\frac{1}{{2m + 1}}\left[ {\exp \left( { - bz} \right)
+ \frac{m}{{m + 1}}\exp \left( { - 2\left( {m + 1} \right)bz}
\right)} \right]~.
\end{align}

Furthermore,  we have~:
\begin{align} \label{Eq:expansion}
&N\sum\limits_{n = 0}^{N - 1} {\left( { - 1} \right)^n }
\binom{N-1}{n}\frac{1}{{2n + 1}}\left[ {\exp \left( { - az} \right)
+ \frac{n}{{n + 1}}\exp \left( { - 2\left( {n + 1} \right)az}
\right)} \right]\\\notag =~&N\sum\limits_{n = 0}^{N - 1} {\left( { -
1} \right)^n } \binom{N-1}{n}\frac{1}{{n + 1}} {\exp \left( { -
2\left( {n + 1} \right)az} \right)} \\\notag +~&N\sum\limits_{n =
0}^{N - 1} {\left( { - 1} \right)^n } \binom{N-1}{n}\frac{1}{{2n +
1}} \left[\exp \left( { - az} \right) - \exp \left( { - 2\left( {n +
1} \right)az} \right)\right]
\\\notag
\buildrel \left(a\right) \over=~&1 - \left[ {1 - \exp \left( { -
2az} \right)} \right]^N  - \exp \left( { - az} \right)\left[
{N\sum\limits_{n = 0}^{N - 1} {\left( { - 1} \right)^n }
\binom{N-1}{n}\frac{1}{{2n + 1}}\left( {1 - \exp \left( { - \left(
{2n + 1} \right)az} \right)} \right)} \right]
\\\notag
\buildrel \left(b\right) \over =~&1 - \left[ { - \sum\limits_{p =
1}^\infty  {\frac{{\left( { - 2az} \right)^p }}{{p!}}} } \right]^N -
\sum\limits_{p = 0}^\infty  {\frac{{\left( { - az} \right)^p
}}{{p!}}} \left[N\sum\limits_{n = 0}^{N - 1} {\left( { - 1}
\right)^n } \binom{N-1}{n}\sum\limits_{p = 1}^\infty  {\left(2n +
1\right)^{p - 1} \frac{{\left( { - az} \right)^p }}{{p!}}}\right]
\\\notag
\buildrel \left(c\right) \over = ~&1 - \frac{1}{2}\left(2az\right)^N
+ o((az)^N )~,
\end{align}
where (a) in (\ref{Eq:expansion}) is achieved by the fact that
$\binom{N-1}{n}\frac{N}{n+1}=\binom{N}{n+1}$ and $\left( {1 - x}
\right)^N = \sum\nolimits_{k = 0}^N {\binom{N}{k}} \left( { - 1}
\right)^k x^k$;~(b)~is achieved by Maclaurian Series of $ \exp
\left( x \right) = \sum\nolimits_{p = 0}^\infty
\left(x^p\right)/\left(p!\right) $; (c) is achieved by
$\sum\nolimits_{k = 0}^N {\left( { - 1} \right)^k \binom{N}{n}k^{n -
1} }  = 0, \left( {1 \le n \le N} \right)$ in
\cite[(0.154.3)]{Gradshteyn94}. Therefore,\begin{align} F_{\gamma
_1} \left(z\right) \approx [\frac{1}{2}\left(2a\right)^N  +
\frac{1}{2}\left(2b\right)^N ]z^N~.\label{Eq:what}
\end{align}
Finally, (\ref{Eq:SER11})~in Theorem 2 is proved by~(\ref{Eq:a}),~
(\ref{Eq:ps2}),~(\ref{Eq:what}),~and $\int\limits_0^\infty {t^{2N} }
\exp \left( -{t^2}/{2}\right)dt={2^{\left(N-{1}/{2}\right)}}\Gamma
\left({1}/{2} + N\right)= \sqrt {{\pi}/{2}}
{{\left(2N\right)!}}/{{\left(2^N N!\right)}} $ in
\cite[(3.326.2)]{Gradshteyn94}, where $\Gamma\left(\cdot\right)$ is
Gamma function\cite{Abramowitz}.

At the case of $\rho_{f_1}<1$ or $\rho_{f_2}<1$, by
$\exp\left(x\right) \approx 1+x$, we similarly have~:
\begin{align}
F_{\gamma _1}  \left( z \right) \approx  &azN\sum\limits_{n = 0}^{N
- 1} {\left( { - 1} \right)^n } \binom{N-1}{n}\frac{{2 - \rho _1
}}{{\left( {2n + 1} \right)\left( {1 - \rho _1 } \right) + 1}}
\\\notag
+ &bzN\sum\limits_{m = 0}^{N - 1} {\left( { - 1} \right)^m }
\binom{N-1}{m}\frac{{2 - \rho _2 }}{{\left( {2m + 1} \right)\left(
{1 - \rho _2 } \right) + 1}}~.
\end{align}
Then, (\ref{Eq:SER12}) in Theorem 2 is proved by (\ref{Eq:a}) and
(\ref{Eq:ps2}).

\newpage

\begin{figure}[h!]
\centering
\includegraphics[height=3.8in,width=4.5in]{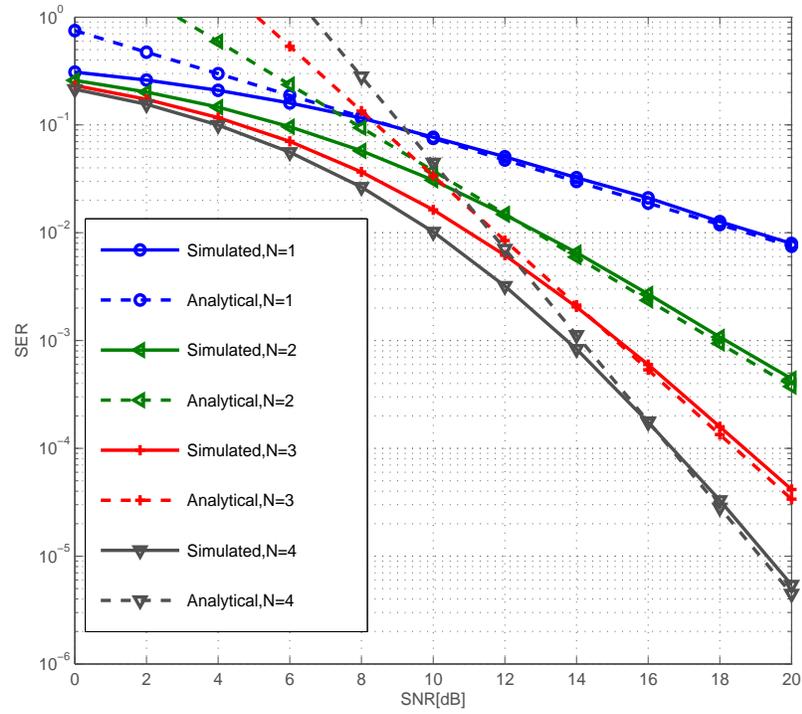}
\caption{Analytical and simulated SER with perfect CSI, with
different $N$, $\rho_{f_1}=\rho_{f_2}=1$ and $\rho_e=1$.}
\label{fig:4}
\end{figure}

\begin{figure}[h!]
\centering
\includegraphics[height=3.8in,width=4.5in]{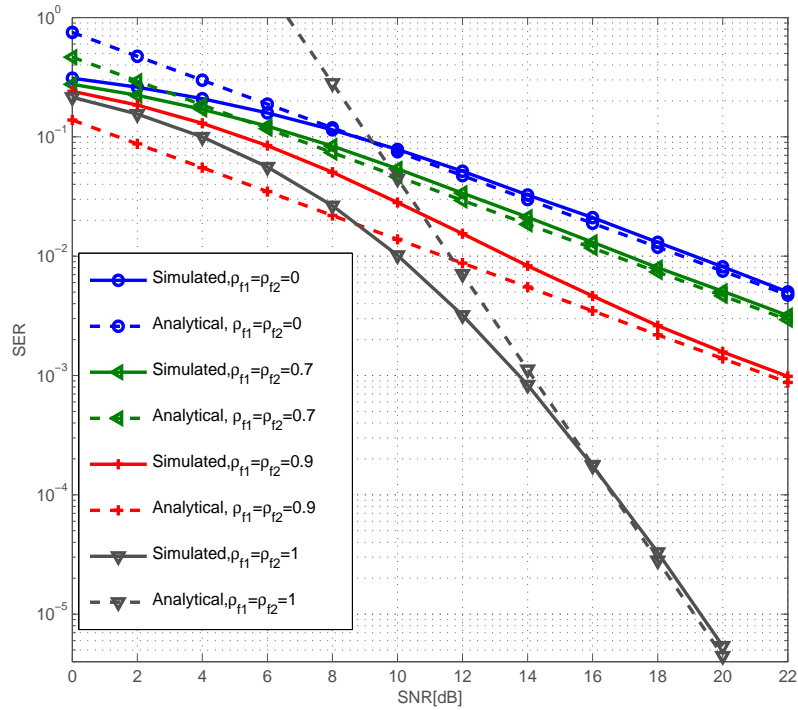}
\caption{Analytical and simulated SER of $S_1$ with delay effect,
with different $\rho_{f_j}$ and $\rho_e=1$, $N=4$.} \label{fig:1}
\end{figure}

\begin{figure}[h!]
\centering
\includegraphics[height=3.8in,width=4.5in]{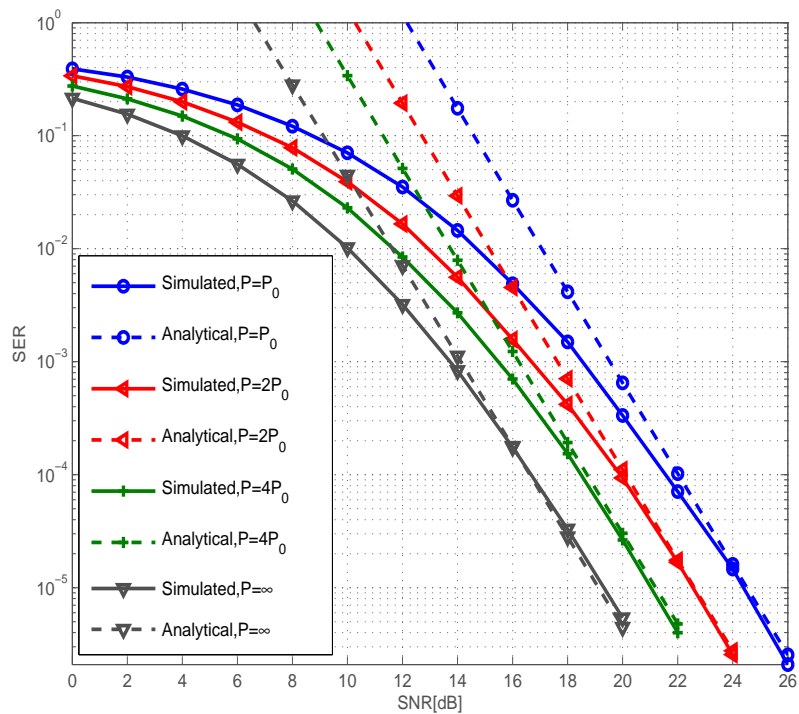}
\caption{Analytical and simulated SER of $S_1$ with estimation
error, with different $\rho_e$ and $\rho_{f_1}=\rho_{f_2}=1$,
$N=4$.} \label{fig:2}
\end{figure}

\begin{figure}[h!]
\centering
\includegraphics[height=3.8in,width=4.5in]{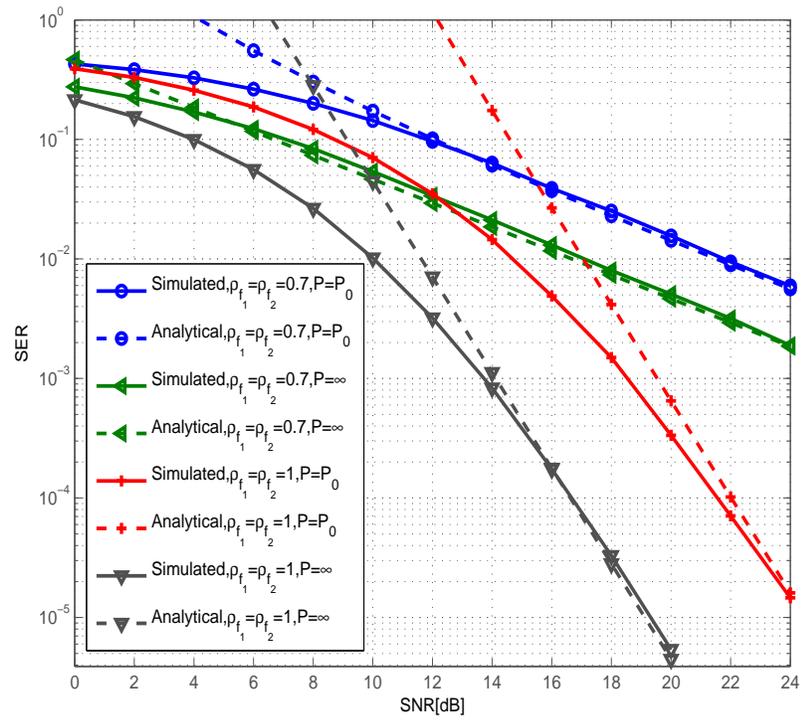}
\caption{Analytical and simulated SER of $S_1$ with delay effect and
estimation error, with different $\rho_e$ and $\rho_{f_i}$, $N=4$.}
\label{fig:3}
\end{figure}

\end{document}